\documentclass[pre,twocolumn,amsmath,amssymb,floatfix,superscriptaddress]{revtex4}

\usepackage{graphicx}
\usepackage{latexsym}
\usepackage{amsmath}
\usepackage{amssymb}
\usepackage{amsfonts}
\usepackage{bm}

\newcommand{\la}{\left<}
\newcommand{\ra}{\right>}

\newcommand{\rvec}{\ensuremath{\underline{r}}}

\newcommand{\vvec}{\ensuremath{\underline{v}}}
\newcommand{\ddiff}{\ensuremath{\text{d}}}
\newcommand{\Mmat}{\ensuremath{\underline{\underline{M}}}}

\newcommand{\rx}{r_x}
\newcommand{\ry}{r_y}
\newcommand{\vx}{v_x}
\newcommand{\vy}{v_y}
\newcommand{\hmat}{\underline{\underline{h}}}

\newcommand{\kBT}{\mbox{$k_{\rm B}T$}}
\newcommand{\NVgT}{\ensuremath{\text{NV}\gamma\text{T}}}
\newcommand{\NVtT}{\ensuremath{\text{NV}\tau\text{T}}}

\newcommand{\Pid}{\ensuremath{P_\mathrm{id}}}
\newcommand{\Pex}{\ensuremath{P_\mathrm{ex}}}

\newcommand{\muA}{\ensuremath{\mu_\mathrm{A}}}
\newcommand{\muAhat}{\ensuremath{\hat{\mu}_\mathrm{A}}}
\newcommand{\muAid}{\ensuremath{\mu_\mathrm{A,id}}}
\newcommand{\muAex}{\ensuremath{\mu_\mathrm{A,ex}}}
\newcommand{\muB}{\ensuremath{\mu_\mathrm{B}}}
\newcommand{\muF}{\ensuremath{\sigma_\mathrm{F}}}
\newcommand{\muFid}{\ensuremath{\sigma_\mathrm{F,id}}}
\newcommand{\muFex}{\ensuremath{\sigma_\mathrm{F,ex}}}
\newcommand{\muFtau}{\ensuremath{\sigma_\mathrm{F}}|_{\tau}}
\newcommand{\muFttau}{\ensuremath{\sigma_\mathrm{F}}(t)|_{\tau}}
\newcommand{\muFgam}{\ensuremath{\sigma_\mathrm{F}}|_{\gamma}}
\newcommand{\muFtgam}{\ensuremath{\sigma_\mathrm{F}}(t)|_{\gamma}}
\newcommand{\muFidgam}{\ensuremath{\sigma_\mathrm{F,id}}|_{\gamma}}
\newcommand{\Cttau}{\ensuremath{C(t)|_{\tau}}}
\newcommand{\Ctgam}{\ensuremath{C(t)|_{\gamma}}}
\newcommand{\Ctgamzero}{\ensuremath{C(0)|_{\gamma}}}
\newcommand{\Cttilde}{\ensuremath{\tilde{C}(t)}}
\newcommand{\Cttautilde}{\ensuremath{\tilde{C}(t)|_{\tau}}}
\newcommand{\Ctgamtilde}{\ensuremath{\tilde{C}(t)|_{\gamma}}}

\newcommand{\Cttaulimit}{C_{\infty}}
\newcommand{\Meq}{M_\mathrm{eq}}
\newcommand{\Geq}{G_\mathrm{eq}}
\newcommand{\GF}{G_\mathrm{F}}
\newcommand{\GFt}{G_\mathrm{F}(t)}

\newcommand{\tauhat}{\hat{\tau}}
\newcommand{\tauidhat}{\hat{\tau}_\mathrm{id}}
\newcommand{\tauexhat}{\hat{\tau}_\mathrm{ex}}

\newcommand{\Ahat}{\ensuremath{\hat{A}}}
\newcommand{\Bhat}{\ensuremath{\hat{B}}}
\newcommand{\dAhat}{\ensuremath{\delta\hat{A}}}
\newcommand{\dBhat}{\ensuremath{\delta\hat{B}}}
\newcommand{\Hhat}{\hat{\cal H}}
\newcommand{\Hidhat}{\hat{\cal H}_\mathrm{id}}
\newcommand{\Hexhat}{\hat{\cal H}_\mathrm{ex}}
\newcommand{\Ihat}{\hat{I}}

\newcommand{\trun}{t_\mathrm{run}}
\newcommand{\gammamax}{\delta \gamma_\mathrm{max}}
\newcommand{\GM}{G_\mathrm{P}}
\newcommand{\tauM}{\tau_\mathrm{P}}
\newcommand{\dtMD}{\delta t_\mathrm{MD}}
\newcommand{\fDebye}{f_\mathrm{Debye}}

\begin{document}

\title{Shear stress relaxation and ensemble transformation\\ 
of shear stress autocorrelation functions revisited}

\author{J.P.~Wittmer}
\email{joachim.wittmer@ics-cnrs.unistra.fr}
\affiliation{Institut Charles Sadron, Universit\'e de Strasbourg \& CNRS, 23 rue du Loess, 67034 Strasbourg Cedex, France}
\author{H.~Xu}
\affiliation{LCP-A2MC, Institut Jean Barriol, Universit\'e de Lorraine \& CNRS, 1 bd Arago, 57078 Metz Cedex 03, France}
\author{J. Baschnagel}
\affiliation{Institut Charles Sadron, Universit\'e de Strasbourg \& CNRS, 23 rue du Loess, 67034 Strasbourg Cedex, France}

\begin{abstract}
We revisit the relation between the shear stress relaxation modulus $G(t)$, computed at finite shear strain 
$0 < \gamma \ll 1$, and the shear stress autocorrelation functions $\Ctgam$ and $\Cttau$
computed, respectively, at imposed strain $\gamma$ and mean stress $\tau$.
Focusing on permanent isotropic spring networks it is shown theoretically and computationally 
that in general $G(t) = \Cttau = \Ctgam + \Geq$
for $t > 0$ with $\Geq$ being the static equilibrium shear modulus. 
$G(t)$ and $\Ctgam$ thus must become different for solids
and it is impossible to obtain $\Geq$ alone from $\Ctgam$ as often assumed. 
We comment briefly on self-assembled transient networks where $\Geq(f)$ must vanish
for a finite scission-recombination frequency $f$. We argue that $G(t) = \Cttau = \Ctgam$ should reveal an 
intermediate plateau set by the shear modulus $\Geq(f=0)$ of the quenched network.
\end{abstract}
\date{\today}
\maketitle

\section{Introduction}
\label{sec_intro}
\begin{figure}[t]
\centerline{\resizebox{1.0\columnwidth}{!}{\includegraphics*{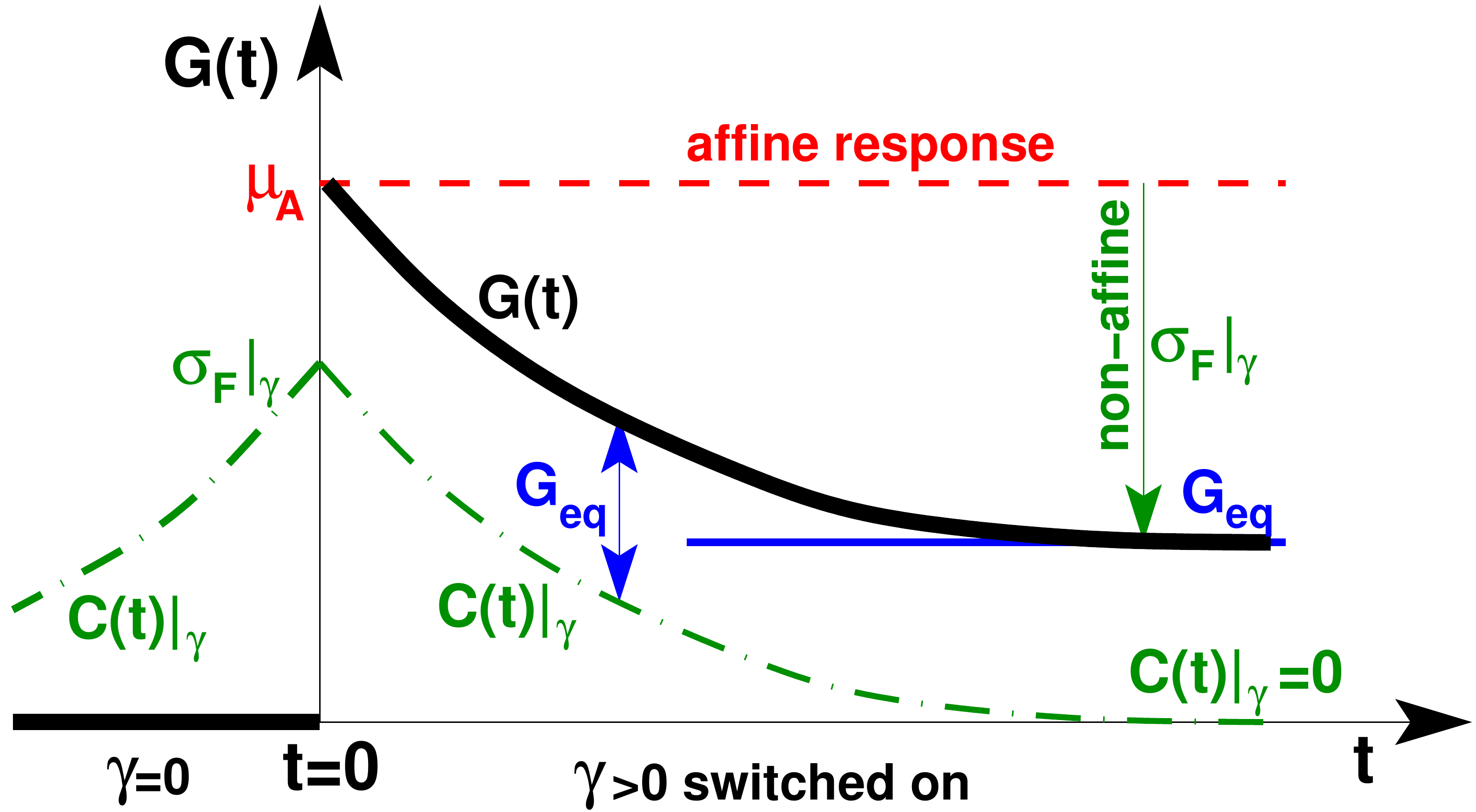}}}
\caption{Schematic comparison of the shear relaxation modulus $G(t)$ (bold solid line) and
the shear stress autocorrelation function $\Ctgam$ computed in the \NVgT-ensemble (dash-dotted line).
Note that $G(0^{+})=\muA = \muFtau$ and $\Ctgamzero = \muFgam$ with $\muA$ being the affine Born-Lam\'e contribution 
to the shear modulus $\Geq = \muA - \muFgam$ 
with $\muF \equiv \beta V \langle \delta \tauhat^2 \rangle$ 
characterizing the static shear stress fluctuations
\cite{Barrat88,Parinello82,Lutsko89,TWLB02,Mezard10,WXP13}.
\label{fig_sketch}
}
\end{figure}

\paragraph*{Shear stress relaxation.}
The static equilibrium shear modulus $\Geq$ 
\cite{LandauElasticity,RubinsteinBook,DoiEdwardsBook,HansenBook,GoetzeBook,Alexander98,WittenPincusBook} 
is an important order parameter \cite{Callen,ChandlerBook,ChaikinBook} characterizing 
the transition from the liquid/sol ($\Geq=0$) to the solid/gel state ($\Geq > 0)$ where the 
particle permutation symmetry of the liquid state is lost for the time window 
probed \cite{Alexander98,WittenPincusBook}.
Examples of current interest for the determination of $\Geq$ include 
crystalline solids \cite{Biroli10},
glass-forming liquids and amorphous solids
\cite{GoetzeBook,Barrat88,WTBL02,TWLB02,Berthier05,Berthier07,Mezard10,Szamel11,SBM11,Yoshino12,Klix12,XWP12,WXP13,WXP13c,ZT13,Ikeda12,Barrat13},
colloidal gels \cite{Kob08},
permanent polymeric networks \cite{RubinsteinBook,DKG91,DKG94,Zippelius06}, 
hyperbranched polymer chains with sticky end-groups \cite{Friedrich10} or
networks of telechelic polymers \cite{Porte03}.
As emphasized by the thin horizontal line in Fig.~\ref{fig_sketch}, the shear modulus of an isotropic
solid may be determined experimentally from the long-time limit \cite{RubinsteinBook,foot_EOS}
\begin{equation}
\Geq \equiv \lim_{t\to \infty} G(t)
\label{eq_Geqdef}
\end{equation}
of the {\em shear stress relaxation modulus} (bold solid line)
defined as $G(t) \equiv \delta \tau(t)/\gamma$. It measures the
stress increment $\delta \tau(t) = \la \tauhat(t) - \tauhat(0^{-}) \ra$  
due to a step strain $0 < \gamma \ll 1$ imposed at time $t=0$. 
Here $\tauhat(t)$ denotes the instantaneous shear stress which may be measured experimentally 
from the forces acting on the walls of the shear cell \cite{RubinsteinBook}.

\paragraph*{Correlation functions.}
A quantity related to $G(t)$ is the {\em shear stress autocorrelation function} 
\cite{HansenBook,GoetzeBook}
\begin{equation}
C(t) \equiv \beta V \la \delta \tauhat(t) \delta \tauhat(0) \ra \equiv \Cttilde - \beta V \langle \tauhat \rangle^2 
\label{eq_Ctdef}
\end{equation}
with $\beta = 1/\kBT$ being  the inverse temperature and $V$ the volume.
We write $\Ctgam$ or $\Cttau$ if $C(t)$ is computed, respectively,
in the \NVgT-ensemble at imposed particle number $N$, volume $V$, shear strain $\gamma$ and 
temperature $T$ or in the conjugated \NVtT-ensemble where instead of $\gamma$ the mean shear 
stress $\tau$ is imposed. 
The effect of the latter constraint is assumed to be arbitrarily slow,
such that $\gamma(t)$ barely changes over the time window probed.
This separation of time scales implies
\begin{equation} 
\Cttautilde = \int \ddiff \gamma \ p(\gamma) \Ctgamtilde
\label{eq_weakbaro}
\end{equation}
with $p(\gamma)$ being the normalized distribution of strains $\gamma$ in the $\NVtT$-ensemble.
The conceptionally important universal limit, Eq.~(\ref{eq_weakbaro}), may be realized
experimentally using an overdamped external force or computationally by either using a 
strong frictional Langevin force added to a standard molecular dynamics (MD) ``shear-barostat" 
\cite{AllenTildesleyBook,FrenkelSmitBook,ThijssenBook,LandauBinderBook,foot_affineshear} 
imposing an average shear stress $\tau$
or (as used below) a Monte Carlo (MC) scheme with a low attempt-frequency for an affine canonical 
$\delta \gamma$-change \cite{WXP13}.

\paragraph*{Key issue.}
Interestingly, it is often assumed \cite{GoetzeBook,DKG91,DKG94,Klix12,AllenTildesleyBook} that 
$G(t)$ and $\Ctgam$ become {\em generally} equivalent in the linear response limit ($\gamma \to 0$). 
If $G(t)=\Ctgam$, the equilibrium shear modulus $\Geq$, Eq.~(\ref{eq_Geqdef}), 
may then be identified with some transient plateau $\GM$ or ``finite frozen-in amplitude" of $\Ctgam$ \cite{Klix12} 
and, hence, with the ``nonergodicity parameter" of the mode-coupling theory for glass-forming liquids
\cite{GoetzeBook}.
Here we raise concerns with such an identification. 
It will be shown that in fact
\begin{equation}
G(t) = \Cttau = \Ctgam + \Geq \mbox{ for } t > 0
\label{eq_key}
\end{equation}
holds for both liquids and solids (and $G(t)=0$ for $t < 0$).
Being implicit to the fluctuation-dissipation theorem (FDT) \cite{DoiEdwardsBook,HansenBook,ChandlerBook,ChaikinBook}
and the general ensemble transformation of dynamical correlation functions
of instantaneous {\em intensive} variables \cite{Green60,Lebowitz67,AllenTildesleyBook},
this is the key relation we want to stress in this paper.
Two important consequences of Eq.~(\ref{eq_key}) are that
{\em (i)}
$G(t)$ only becomes equivalent to $\Ctgam$ for $t > 0$ in the liquid limit where $\Geq=0$ and that 
{\em (ii)}
a {\em finite} shear modulus $\Geq$ is only probed by $G(t)$ on time scales where $\Ctgam$ actually vanishes. 
While the static shear modulus $\Geq$ can be obtained from $\Cttau$, this is not possible 
using only $\Ctgam$ without making additional model-specific assumptions.

\paragraph*{Outline.}
We recall in Sec.~\ref{theo_static} the ``affine" contribution $\muA$ and the 
``stress fluctuation" contribution $\muFgam$ to the equilibrium shear modulus $\Geq = \muA - \muFgam$.
The key relation Eq.~(\ref{eq_key}) is then demonstrated theoretically in Sec.~\ref{theo_dyna}
using several (albeit not completely independent) lines of thought.
If the stress fluctuation contribution $\muF(t)$ is determined numerically over a finite time
window $t$, it must systematically underestimate the value $\muF$ for asymptotically long
sampling times \cite{foot_tdependence}. It is seen in Sec.~\ref{theo_muFtCt} how $\muF(t)$
is quite generally related to the correlation function $C(t)$. 
We briefly comment on self-assembled transient elastic networks in Sec.~\ref{theo_trans}.
The specific model system considered numerically is introduced in Sec.~\ref{sec_model}.
A well-defined solid with finite equilibrium shear modulus $\Geq$ for $t \to \infty$ is assumed.
For this reason we replace the Lennard-Jones (LJ) interactions of a quenched bead system by a {\em permanent} 
elastic spring network corresponding to its dynamical matrix at zero temperature 
\cite{WTBL02,TWLB02,WXP13}.
Some static properties and measurement procedures are summarized in Sec.~\ref{simu_static}.
Using our simple model Hamiltonian the key relation is confirmed numerically in Sec.~\ref{simu_dyna} 
by means of molecular dynamics (MD), 
Brownian dynamics (BD) and Monte Carlo (MC) simulations 
\cite{AllenTildesleyBook,FrenkelSmitBook,ThijssenBook,LandauBinderBook}.
This work is summarized in Sec.~\ref{sec_conc}.
We finally state the generalization of Eq.~(\ref{eq_key}) for autocorrelation functions of other intensive variables
and comment briefly on ongoing simulations of self-assembled transient networks. 

\section{Theoretical considerations}
\label{sec_theo}

\subsection{Static properties}
\label{theo_static}

\paragraph*{Static stress fluctuations.}
We begin by reminding \cite{WXP13} that the shear modulus $\Geq$ of a solid body may be obtained 
in principle from
\begin{equation}
\muFtau = \muFgam + \Geq
\label{eq_muFtaumuFgam}
\end{equation}
by comparing the (reduced) shear stress fluctuations 
\begin{equation}
\muF \equiv C(t=0) \equiv \beta V \la \delta \tauhat^2 \ra
\label{eq_muFdef}
\end{equation}
at constant mean shear stress $\tau$ (\NVtT-ensemble) with the fluctuations at imposed strain $\gamma$ (\NVgT-ensemble).
This relation is obtained directly from the Lebowitz-Percus-Verlet transformation for a fluctuation 
$\langle \dAhat \dBhat \rangle$ of two observables ${\cal A}$ and ${\cal B}$ 
\cite{Lebowitz67,AllenTildesleyBook,WXP13}
\begin{equation}
\left. \la \dAhat \dBhat \ra\right|_{I} =
\left. \la \dAhat \dBhat \ra\right|_{X}  
+ \frac{\partial (\beta I)}{\partial X} \frac{\partial \langle \Ahat \rangle}{\partial (\beta I)} 
\frac{\partial \langle \Bhat \rangle}{\partial (\beta I)}
\label{eq_dAdB}
\end{equation}
with $X = V\gamma$ being in our case the extensive variable, $I = \tau$ the conjugated intensive variable
and $\Ahat = \Bhat = \tauhat$ \cite{Callen}.
For the simplicity of the notation we have assumed in Eq.~(\ref{eq_dAdB}) 
that $X$ is {\em not} the internal energy $U$.
For a more general theoretical description it is necessary to define the
``entropic intensive variable" $J \equiv \partial S(X) / \partial X$ with $S(X)$ being the
entropy \cite{Callen}.  If $X \ne U$, one has $J = - I/T$ \cite{Callen}.
These entropic intensive variables are used in Ref.~\cite{Lebowitz67}.
Note that expressing Eq.~(\ref{eq_dAdB}) in terms of $I$, rather than in terms of $J$, changes the signs.

\paragraph*{From fluctuations to simple means.}
From the computational point of view it is important that Eq.~(\ref{eq_muFtaumuFgam}) can be further simplified.
With $\Hhat(\gamma) = \Hidhat(\gamma) + \Hexhat(\gamma)$ being the Hamiltonian of a given state 
of the system parameterized in terms of an affine strain $\gamma$ 
\cite{Parinello82,Lutsko89,WXP13,foot_affineshear},
its normalized weight in the \NVtT-ensemble is given by $p(\gamma) \sim \exp[-\beta (\Hhat(\gamma) -V\gamma \tau)]$.
We thus have
\begin{equation}
p^{\prime}(\gamma) = - \beta V [\tauhat(\gamma) - \tau)] p(\gamma)
\mbox{ with } \tauhat(\gamma) \equiv \Hhat^{\prime}(\gamma)/V
\label{eq_pprime}
\end{equation}
{\em defining} the instantaneous shear stress \cite{WXP13}. 
(A prime denotes a derivative of a function with respect to its argument.)
For small $\gamma$ it follows that $\tauidhat \equiv \Hidhat^{\prime}(\gamma)/V$ reduces to 
the standard instantaneous ideal shear stress and $\tauexhat \equiv \Hexhat^{\prime}(\gamma)/V$ 
for pair potential interactions to the Kirkwood virial expression 
of the shear stress \cite{AllenTildesleyBook,WXP13,foot_Ihat}. 
By integration by parts the stress fluctuation $\muFtau$
can be expressed as the ``simple average" \cite{WXP13,WXP13c}
\begin{equation}
\muFtau = \frac{1}{V}\la \Hhat^{\prime\prime}(\gamma)\ra
= \la \tauhat^{\prime}(\gamma) \ra|_{\tau} \equiv \muA
\label{eq_muA}
\end{equation}
which can be directly computed in any ensemble assuming that the same state point is sampled.
The ``affine shear elasticity" $\muA$ characterizes the mean total (kinetic and excess) 
energy change $\muA V \gamma^2/2$ assuming a {\em homogeneous affine} shear transformation of the system 
as it may be done in a computer experiment by changing the metric of system
\cite{AllenTildesleyBook,FrenkelSmitBook,Parinello82,foot_affineshear}.
For pair potentials $\muA$ can be further reduced to 
\begin{equation}
\muA = \muB - \Pex + \Pid
\label{eq_muAmuB}
\end{equation}
with $\muB$ being the well-known Born-Lam\'e coefficient, $\Pex$ the excess pressure
and $\Pid$ the ideal pressure contribution. We have thus rewritten $\muFtau$ 
as a simple average of moments of first and second derivatives of the potential plus $\Pid$. 
(Since second derivatives are considered, impulsive corrections must be taken
into account for truncated and shifted potentials as stressed in Ref.~\cite{XWP12}.)
The shear modulus can hence be conveniently computed by means of the stress-fluctuation formula 
\begin{equation}
\Geq = \GF \equiv \muA - \muFgam
\label{eq_GeqNVgT}
\end{equation}
in the $\NVgT$-ensemble \cite{Barrat88,Lutsko89,TWLB02,Mezard10,WXP13,Barrat13}.
Since for a plain shear strain at constant volume the ideal free energy
contribution does not change, the explicit kinetic energy contributions
must be irrelevant for $\Geq$. (An ideal gas cannot elastically support a finite shear stress.)
As one thus expects, the kinetic contributions $\muAid = \muFidgam = \Pid$ to $\muA$ and $\muFgam$ 
cancel and can be dropped when $\Geq$ is determined using Eq.~(\ref{eq_GeqNVgT}).

\subsection{Demonstration of key relation}
\label{theo_dyna}

\paragraph*{Asymptotic limits.}
As shown by the dash-dotted line in Fig.~\ref{fig_sketch}, by definition $\Ctgam \to \muFgam$ 
for $t \to 0$ and $\Ctgam \to 0$ for $t \to \infty$ \cite{HansenBook}. 
Equation~(\ref{eq_key}) thus implies that $G(t) \to \muFgam + (\muA-\muFgam) = \muA$ for $t \to 0^{+}$ 
--- which is consistent with the affine shear strain imposed at $t=0$ --- 
and $G(t) \to \Geq$ for $t \to \infty$ as it should.
We note also that by definition $\Cttau \to \muFtau = \muA$ for $t \to 0$.
Interestingly, the autocorrelation function $\Cttau$ does not vanish in general in the large-$t$ limit.
This is a direct consequence of the time scale separation mentioned above, Eq.~(\ref{eq_weakbaro}), 
from which it is seen that
\begin{eqnarray}
\Cttau & = &  \int \ddiff \gamma  \ p(\gamma) \Ctgam + \Cttaulimit \mbox{ with }
\label{eq_CttauCtgamOne} \\
\Cttaulimit & \equiv & \beta V \int \ddiff \gamma \ p(\gamma) \left( \underline{\langle \tauhat \rangle|_{\gamma}^2} -\tau^2 \right).
\label{eq_CttauCtgamTwo}
\end{eqnarray}
The first contribution to $\Cttau$ in Eq.~(\ref{eq_CttauCtgamOne}) vanishes for $t \to \infty$.
Note that $\Cttaulimit$ differs from $\muFtau = \muA$ due to the underlined term in Eq.~(\ref{eq_CttauCtgamTwo}).
Using that $p(\gamma)$ is Gaussian and $\langle \delta \gamma^2 \rangle = \kBT /(V\Geq)$, it is seen that 
\begin{equation}
\Cttau \to \Cttaulimit = \beta V \Geq^2 \la \delta \gamma^2 \ra = \Geq \mbox{ for } t \to \infty.
\label{eq_Ctaulimit}
\end{equation}
We show now that Eq.~(\ref{eq_key}) must hold for all times.

\paragraph*{First equality of the key relation.}
Generalizing Eq.~(\ref{eq_muA}) one shows for the shear stress fluctuations at constant stress 
(assuming a slow shear-barostat) that
\begin{equation}
\Cttau = \left. \la \frac{\partial \tauhat(t;\gamma)}{\partial \gamma} \ra \right|_{\tau} = G(t) \mbox{ for } t > 0.
\label{eq_GtCttau}
\end{equation}
To show this, we have reexpressed in the first step $[\tauhat(t;\gamma)-\tau] [\tauhat(0;\gamma)-\tau] p(\gamma)$ 
using Eq.~(\ref{eq_pprime}) and integration by parts. In the second step 
we have used that within linear response $G(t)$ does not depend on $\gamma$.
This demonstrates the first equality stated in Eq.~(\ref{eq_key}).
\paragraph*{Second equality of the key relation.}
Using Boltzmann's superposition principle the shear stress $\tau(t)$ for an arbitrary 
strain history $\gamma(t)$ may be written \cite{RubinsteinBook,DoiEdwardsBook}
\begin{eqnarray}
\tau(t) & = & \int_{-\infty}^{t} \ddiff s \ G(t-s) \frac{d \gamma(s)}{d s} \label{eq_Boltzsup1} \\
        & = & \left. G(t-s) \gamma(s) \right|_{-\infty}^{t} - 
\int_{-\infty}^{t} \ddiff s \frac{d G(t-s)}{d s} \gamma(s) \nonumber 
\end{eqnarray} 
using integration by parts. Since $\gamma(t)$ is a step function 
and introducing the ``after-effect function" $\chi(t) \equiv - G^{\prime}(t) = G^{\prime}(-t)$ 
\cite{HansenBook} this gives
\begin{equation}
G(t) = G(0) - \int_0^t \chi(s) \ \ddiff s =
\Geq + \int_t^{\infty} \chi(s) \ \ddiff s 
\label{eq_Gtchit}
\end{equation}
where $\Geq$ appears as an integration constant.
Since according to the FDT as formulated by Eq.~(7.6.13) of Ref.~\cite{HansenBook},
the after-effect function is given by $\chi(t) = - C^{\prime}(t)|_{\gamma}$,
this demonstrates $G(t) = \Geq + \Ctgam$ 
as stated by the second equality in Eq.~(\ref{eq_key}) \cite{foot_Edwards}.
Alternatively, from Eq.~(\ref{eq_CttauCtgamOne}) and Eq.~(\ref{eq_Ctaulimit}) one obtains directly
\begin{equation}
\Cttau \to \Ctgam + \Geq \mbox{ for } V \to \infty
\label{eq_Cttauctgam}
\end{equation}
using steepest-descent, $\int \ddiff \gamma  \ p(\gamma) \Ctgam \to \Ctgam$, 
with $\gamma$ corresponding to the maximum of $p(\gamma)$. Together with Eq.~(\ref{eq_GtCttau})
this confirms again our key relation.

\paragraph*{Dynamical Lebowitz-Percus-Verlet transform.}
Interestingly, Eq.~(\ref{eq_Cttauctgam}) may be also obtained by generalizing
the Lebowitz-Percus-Verlet transformation, Eq.~(\ref{eq_dAdB}), into the time domain with 
$\Ahat = \tauhat(t)$ and $\Bhat = \tauhat(0)$.
We remind that this transform relies on the condition 
that only the distribution of start points of the trajectories depends on the ensemble, 
but not the relaxation pathways themselves \cite{AllenTildesleyBook}. While this does not 
hold for extensive variables if the {\em same} extensive variable if imposed, 
this is generally the case for fluctuations of instantaneous intensive variables
which we focus on here.
Interestingly, a similar approach based on Ref.~\cite{Lebowitz67} has been used for the 
four-point dynamic susceptibility $\chi_4(t)$ comparing its decay at constant temperature 
and constant energy \cite{Berthier05,Berthier07}.

\subsection{Time dependence of stress fluctuations}
\label{theo_muFtCt}

\paragraph*{Introduction.}
We have seen in Sec.~\ref{theo_static} that the shear modulus $\Geq$ may be obtained by 
measuring the static stress fluctuations $\muF = \beta V \langle \delta \tauhat^2 \rangle$.
As for any fluctuation measured along a trajectory \cite{AllenTildesleyBook,LandauBinderBook} 
one expects the stress fluctuations $\muF(t)$ computed over a too short time window $t$ to yield 
only a time-dependent {\em lower bound} to the true asymptotic long-time limit $\muF$ \cite{foot_tdependence}.
(This remains even true if as a second step one averages over independent trajectories.)
This may seriously restrict the use of the stress-fluctuation formula, Eq.~(\ref{eq_GeqNVgT}),
as will be seen at the end of Sec.~\ref{simu_static} below.
It is thus important that for systems with time translational symmetry
$\muF(t)$ can be rewritten as an integral over the stress autocorrelation function $C(t)$. 

\paragraph*{Correlated trajectories.}
Let us consider $N \gg 1$ successive observations $x_n \equiv \sqrt{\beta V} \tauhat(n)$
with $n=1,\ldots, N$ stored at equidistant time steps $\delta t$ over a
total time interval $t = N \delta t$. Using similar steps as for the calculation of the radius of gyration of
polymer chains (Sec. 2.4 of Ref.~\cite{DoiEdwardsBook}) one may rewrite the expectation value 
$\langle (x_n - \langle x_n \rangle)^2\rangle$ of the 
shear stress fluctuations as
\begin{equation}
\muF(N) = \la \frac{1}{2N^2} \sum_{n,m=1}^N \left( x_n - x_m  \right)^2 \ra
\label{eq_muFt1}
\end{equation}
where the average is performed over different trajectories.
Defining a ``mean-square displacement" $g(s) \equiv \langle (x_{m=n+s} - x_n)^2 \rangle$ 
this allows to rewrite Eq.~(\ref{eq_muFt1}) as
\begin{equation}
\muF(N) = \frac{1}{N^2} \sum_{s=1}^N (N-s) g(s).
\label{eq_muFt2}
\end{equation}
where the weight $(N-s)$ stems from the finite trajectory length. 
Using the correlation function $C(s) = \langle x_{m=n+s} x_n \rangle$ 
one verifies that $g(s) = 2 (C(0)-C(s))$ \cite{DoiEdwardsBook}.
Since $\sum_{s=1}^N (N-s) \approx N^2/2$ to leading order, this implies in turn
\begin{equation}
\muF(N) = C(0) - \frac{2}{N^2} \sum_{s=1}^{N} (N - s) C(s). 
\label{eq_muFt3}
\end{equation}
Using that $C(0) = \muF$ and rewriting the discrete sum as a continuous
time integral this yields
\begin{equation}
\muF(t) = \muF - \frac{2}{t} \int_0^t \ddiff s \ (1 - s/t) C(s)
\label{eq_muFtCt}
\end{equation}
independent of whether $\gamma$ or $\tau$ are imposed.
It follows that the stress-fluctuation formula $\GF \equiv \muA - \muFgam$ 
may be rewritten quite generally as \cite{foot_tdependence}
\begin{eqnarray}
\GFt & =& \Geq + \frac{2}{t} \int_0^t \ddiff s \ (1 - s/t) \ C(s)|_{\gamma} \label{eq_GFtCt} \\
& = & \frac{2}{t} \int_0^t \ddiff s \ (1 - s/t) \ G(s) \label{eq_GFtGt}
\end{eqnarray}
where we have used the key relation, Eq.~(\ref{eq_key}), in the second step.
For large times $\Ctgam \to 0$ and the integral over $\Ctgam$ becomes constant.
As expected for general finite-sampling time corrections for fluctuations \cite{LandauBinderBook},
the second term in Eq.~(\ref{eq_GFtCt}) vanishes thus extremely slowly as $1/t$ \cite{foot_Debye}.
As seen from Eq.~(\ref{eq_GFtGt}), $\GFt$ and $G(t)$ are different in general albeit
closely related. They have the same asymptotic limits $\GF(0) = G(0) = \muA$ and $\GF(\infty)=G(\infty) = \Geq$. 

\paragraph*{Intermediate plateau.}
It is of some interest to consider briefly the case of model systems where
the stress autocorrelation function $\Ctgam$ reveals a broad intermediate plateau, 
$\Ctgam = \GM$, extending over several orders of magnitude up to a time $\tauM$. 
It is readily seen using Eq.~(\ref{eq_GFtCt}) or Eq.~(\ref{eq_GFtGt}) that 
\begin{equation}
G(t) = \Cttau \approx \Geq + \GM \approx \GFt \mbox{ for } t \ll \tauM,
\label{eq_GtGFtGM}
\end{equation}
i.e. $G(t)$ and $\GFt$ may become identical and constant for a finite time window.

\subsection{Digression: Self-assembled transient networks}
\label{theo_trans}
While the present work focuses on {\em permanent} elastic networks 
let us mention that this study can be extended naturally on self-assembled transient networks as 
hyperbranched polymer chains with sticky end-groups \cite{Friedrich10}
or microemulsions bridged by telechelic polymers \cite{Porte03}.
Such networks may be modeled using purely repulsive LJ beads representing
the oil droplets of the microemulsion which are connected reversibly by ideal springs
similarly as in MC simulations of equilibrium polymers \cite{WMC98b}.
The topological rearrangement of the network may be done by randomly choosing
a spring with a scission-recombination frequency $f$ and making a hopping attempt 
to reconnect it with other neighboring beads subject to a standard Metropolis 
criterion \cite{LandauBinderBook}.
If one freezes an equilibrated network, i.e. if one sets $f=0$,
the network must behave exactly as the permanent solids we focus on
in this work, i.e. Eq.~(\ref{eq_key}) should hold with a finite $\Geq(f=0)$.
If one considers a very small, but finite frequency $f$ and very long sampling times,
one expects $\GFt$ to show an intermediate plateau $\GM$ up to the relaxation time of 
the network $\tauM(f) \sim 1/f$ and to decay for larger times. 
($\tauM(f)$ characterizes the time needed to restore the particle permutation
symmetry of the liquid state.)
The plateau $\GM$ of $G(t)$ should be set by the modulus $\Geq(f=0)$ of the quenched network, 
since for small times $t \ll \tauM(f)$ the scission-recombination events become irrelevant.
Since $\Geq(f) = 0$ for finite $f$, this implies according to Eq.~(\ref{eq_key})
that $G(t) = \Cttau = \Ctgam$ for all times.
Using now Eq.~(\ref{eq_GFtCt}) and Eq.~(\ref{eq_GtGFtGM}) these arguments suggest
confirming Ref.~\cite{Biroli10} that 
\begin{equation}
G(t) = \Ctgam = \GFt = \GM = \Geq(f=0)
\label{eq_transnet}
\end{equation}
for intermediate times $t \ll \tauM(f)$.
In this sense $\Ctgam$ may indeed measure a shear modulus. It is however
{\em not} the shear modulus $\Geq(f)$ of the system computed at finite $f$
(which must vanish) but of the quenched reference network at $f=0$.
We return after this digression to solids formed by permanently connected springs.

\section{Computational model}
\label{sec_model}
To illustrate our key relation we present in Sec.~\ref{sec_simu} numerical data obtained 
using a periodic two-dimensional network of ideal harmonic springs of interaction energy 
\begin{equation}
\Hexhat = \frac{1}{2} \sum_l K_l \left(r_l - R_l\right)^2
\label{eq_Enet}
\end{equation}
with $K_l$ being the spring constant, $R_l$ the reference length and $r_l = |\rvec_i - \rvec_j|$ 
the length of spring $l$.
The sum runs over all springs $l$ between topologically connected vertices $i$ and $j$ 
of the network at positions $\rvec_i$ and $\rvec_j$.
Note that the mass of the particles is set to unity and LJ units are assumed throughout. 
As explained in detail in Ref.~\cite{WXP13}, our network has been constructed using the 
dynamical matrix $\Mmat$ of a strongly polydisperse LJ bead glass comprising $N=10^4$ particles. 
(An experimentally more relevant example for such permanent networks is provided by 
endlinked or vulcanized polymer networks \cite{RubinsteinBook,DKG91,DKG94}.)
Prior to forming the network the latter bead system had been quenched down to $T=0$ using a constant 
quenching rate and imposing a relatively large normal pressure $P=2$.
This yields systems of number density $\rho \approx 0.96$, 
i.e.  linear periodic box length $L \approx 102.3$.
Since the network topology is by construction {\em permanently fixed}, the shear response $G(t)$ 
must become finite for $t \to \infty$ for all temperatures at variance to systems with plastic 
rearrangements as considered, e.g., in Ref.~\cite{Biroli10}, or the transient networks
mentioned in Sec.~\ref{theo_trans}.
\section{Computational results}
\label{sec_simu}

\subsection{Static properties}
\label{simu_static}

\paragraph*{Ground state values.}
Following Refs.~\cite{WTBL02,TWLB02} one may compute the shear modulus $\Geq$ of the ground state
of the model at $T=0$ from the lowest non-trivial four-fold degenerated eigenfrequencies $\omega_p$
associated to transverse eigenmodes. (The running index $p$ increases with frequency.)
Such eigenmodes can be determined by numerical diagonalization 
of the dynamical matrix $\Mmat$ by means of Lanczos' method \cite{ThijssenBook}.  
For planar transverse modes one expects from continuum theory \cite{LandauElasticity} that
\begin{equation}
\omega_p = \frac{2\pi c_T}{L} \sqrt{n^2 + m^2} \mbox{ with } c_T = \sqrt{\Geq/\rho}
\label{eq_omegap}
\end{equation}
being the transverse wave velocity and $n,m=0,1,\ldots$ two quantum numbers.
One thus obtains for $p=3,4,5,6$ that $\Geq \approx 16$.
By applying an affine shear strain to the system or by using Eq.~(\ref{eq_muA})
one determines an affine shear elasticity $\muA \approx 34$. In turn this implies 
a shear stress fluctuation $\muFgam = \muA - \Geq \approx 18$, i.e. about half
of the energy implied by an affine strain is relaxed by non-affine displacements 
as discussed in more detail in Ref.~\cite{TWLB02}.
%
%

\begin{figure}[t]
\centerline{\resizebox{1.0\columnwidth}{!}{\includegraphics*{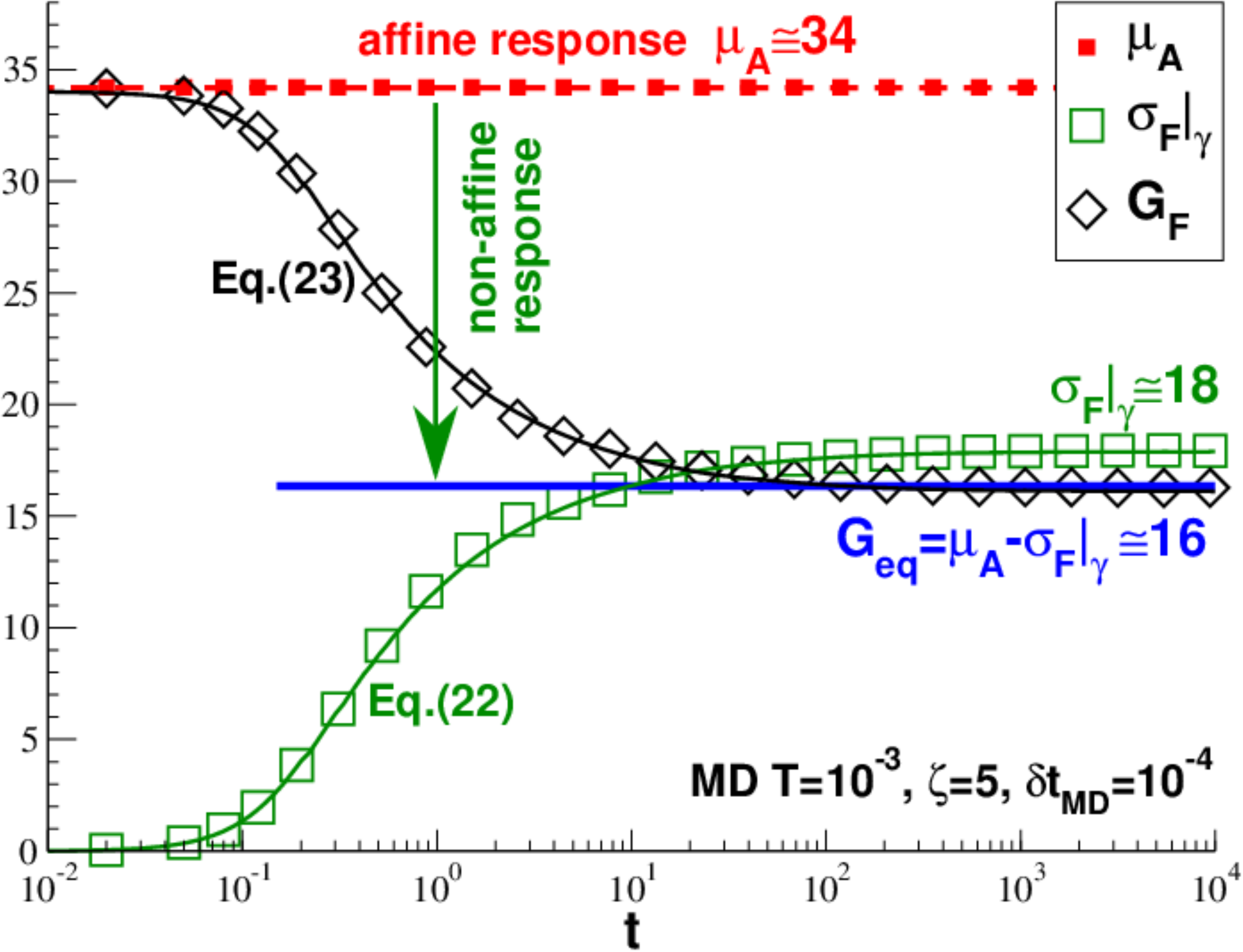}}}
\caption{Affine shear elasticity $\muA(t)$, shear stress fluctuations $\muFtgam$
and their difference $\GFt = \muA- \muFtgam$ as a function of the measurement
time $t$ for one network of ideal springs in two dimensions (\NVgT-ensemble).
The data have been sampled by MD simulation with time step $\dtMD = 10^{-4}$,
temperature $T=10^{-3}$ and friction constant $\zeta=5$. 
The solid lines present a consistency check of $\muFtgam$ and $\GFt$
with the correlation function $\Ctgam$ confirming 
Eq.~(\ref{eq_muFtCt}) and Eq.~(\ref{eq_GFtCt}).
\label{fig_muAmuFgam}
}
\end{figure}

\paragraph*{Static properties at small finite temperatures.}
We focus below on systems with a finite temperature $T = 10^{-3}$.
Since this temperature is rather small, one expects all static properties
such as  the pressure $P$ or the elastic modulus $\Geq$ to be barely changed.
As we have checked comparing various methods one confirms indeed that 
$P \approx \Pex \approx 2$, $\muA \approx 34$, $\muFgam \approx 18$ and $\Geq \approx 16$
and the same applies to all small temperatures $T \ll 1$. 

\paragraph*{Convergence of stress-fluctuation formula.}
How the shear modulus is obtained using the stress-fluctuating formula, Eq.~(\ref{eq_GeqNVgT}),
can be seen from Fig.~\ref{fig_muAmuFgam} where we present data obtained by 
standard velocity-Verlet MD simulation \cite{AllenTildesleyBook,FrenkelSmitBook} 
using a (rather cautious) time step $\dtMD = 10^{-4}$. 
The temperature $T=10^{-3}$ is imposed by means of a Langevin thermostat with a large friction 
constant $\zeta=5$. We have used here one long production run over a time $\trun = 10^5$
for one equilibrated start configuration. Various properties, such as the instantaneous values of the
shear stress $\tauhat(t)$ or the affine shear elasticity $\muAhat(t)$, Eq.~(\ref{eq_muAmuB}),
have been written down at equidistant time steps $\delta t = 10^{-2}$.
The data correspond to averages taken first over a given time interval 
$[t_0,t_1=t_0+t]$, i.e. using $1+t/\delta t$ entries, and taking then in a second step 
gliding averages over all times $t_0$ possible for $t$ \cite{AllenTildesleyBook}. 
(Naturally, the error bars thus increase somewhat with $t$.)
The horizontal axis in Fig.~\ref{fig_muAmuFgam} indicates the interval length $t$.
As one expects, the simple average $\muA(t)$ becomes immediately constant (filled squares), 
i.e. $\muA(t) = \muA \approx 34$, as indicated by the dashed horizontal line \cite{foot_tdependence}. 
By contrast, the shear stress fluctuations $\muFtgam$ are seen 
to increase monotonously from zero to the asymptotic plateau $\muFgam \approx 18$. 
Interestingly, this plateau is only reached for surprisingly large times $t \gg 10^3$. 
The stress-fluctuation estimate $\GFt \equiv \muA - \muFtgam$ (diamonds) of the shear modulus $\Geq$
decreases thus monotonously from $\muA$ to its large-$t$ limit $\Geq \approx 16$ indicated
by the bold horizontal line. A too short production run thus leads to an overestimation of $\Geq$.
The two thin solid lines present a consistency check for $\muFtgam$ and $\GFt$
integrating the shear stress autocorrelation function $\Ctgam$ as suggested
by Eq.~(\ref{eq_muFtCt}) and Eq.~(\ref{eq_GFtCt}). The slow convergence 
of the stress-fluctuation relation $\GFt$ noted in Refs.~\cite{SBM11,WXP13} can 
thus be traced back to the sluggish $1/t$-decay of the second term in Eq.~(\ref{eq_GFtCt}).
We turn now to the description of $\Ctgam$ and other correlation functions.

\subsection{Computational test of key relation}
\label{simu_dyna}

\begin{figure}[t]
\centerline{\resizebox{1.0\columnwidth}{!}{\includegraphics*{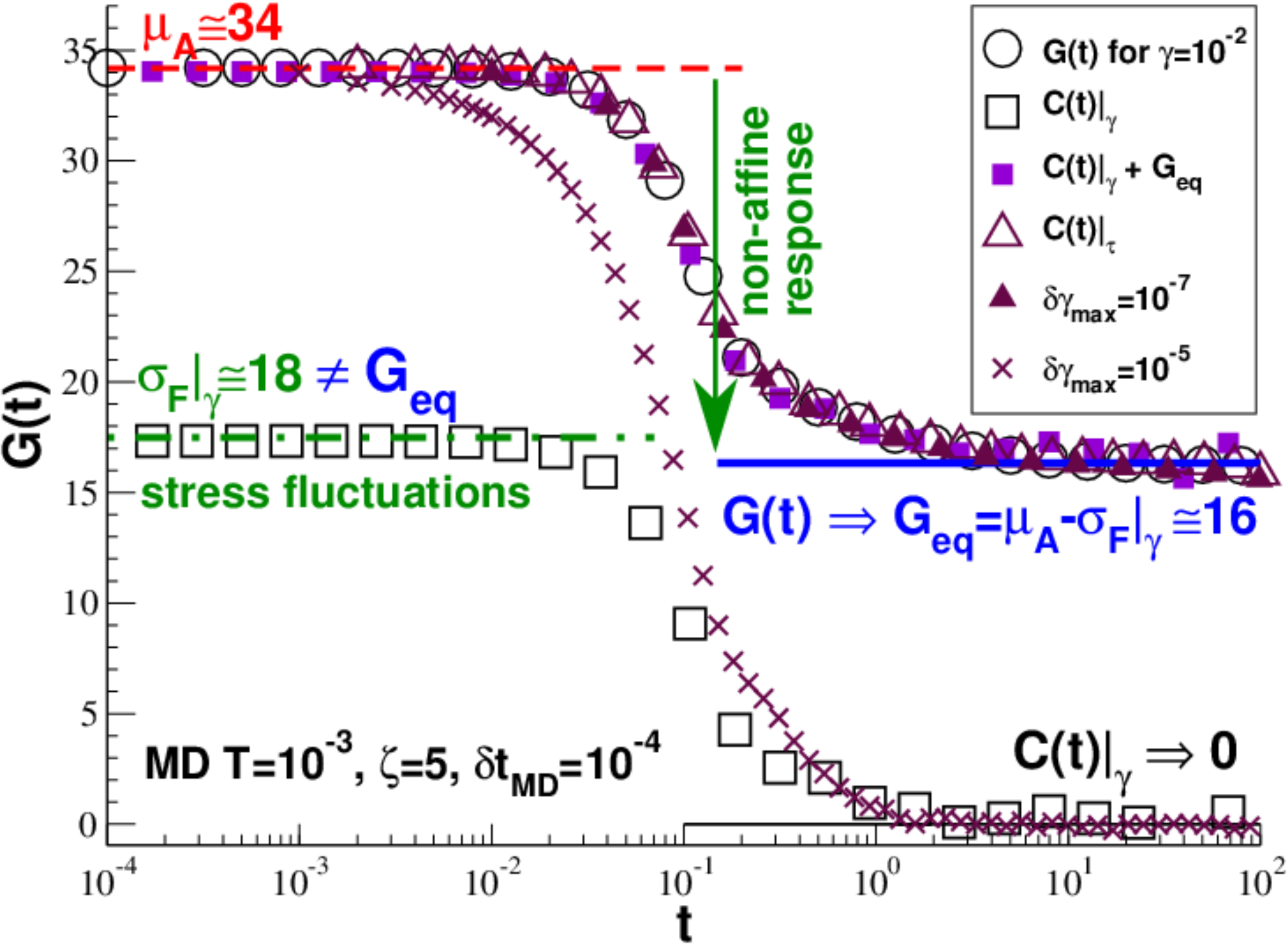}}}
\caption{Stress relaxation modulus $G(t)$ and stress autocorrelation functions
$\Cttau$ (triangles) and  $\Ctgam$ (squares) sampled by MD simulation.
An affine strain $\gamma=10^{-2}$ is applied to determine $G(t)$.
The filled triangles correspond to $\Cttau$ computed using one single
trajectory with a slow MC shear-barostat with $\gammamax=10^{-7}$,
the crosses to a too large value $\gammamax=10^{-5}$ 
for which $\Cttau$ is seen to decay rapidly
due to additional relaxation pathways. 
\label{fig_dataMD}
}
\end{figure}

\paragraph*{Introduction.}
Having shown in Fig.~\ref{fig_muAmuFgam} how a finite shear modulus $\Geq \approx 16$ may
be determined in the \NVgT-ensemble using the stress fluctuation formula, Eq.~(\ref{eq_GeqNVgT}),
we now demonstrate numerically our key relation, Eq.~(\ref{eq_key}), by comparing the explicitly
computed out-of-equilibrium stress relaxation modulus $G(t)$ with the equilibrium autocorrelation 
functions $\Ctgam$ and $\Cttau$. 
As before we show first in Fig.~\ref{fig_dataMD} data obtained by MD simulations using a high 
friction constant $\zeta=5$, which simplifies the data by enforcing a monotonic decay of the correlations.
We discuss then results obtained using different friction constants and computational schemes.

\paragraph*{Stress relaxation and autocorrelation functions.}
The stress relaxation modulus $G(t)$ presented in Fig.~\ref{fig_dataMD} has been computed 
from the shear stress increment $\delta \tauhat(t)$ measured after an affine shear strain $\gamma=10^{-2}$ 
was imposed at $t=0$ \cite{foot_affineshear}.
We average over $10^3$ runs starting from independent reference configurations at $t=0^{-}$.
The shear stress relaxation modulus $G(t)$ decreases (due to the strong damping) monotonously 
from $G(0^{+})=\muA$ to a finite $\Geq$. In contrast to this $\Ctgam$ (open squares) decays from $\muFgam$ to zero. 
Confirming Eq.~(\ref{eq_key}), the vertically shifted autocorrelation function 
$\Ctgam+\Geq$ (filled squares) is seen to collapse onto $G(t)$. 
The autocorrelation function $\Cttau$ (open triangles) has been obtained by preparing 
first an \NVtT-ensemble of mean stress $\tau=0$ containing $10^4$ independent start configurations. 
We sample $\Ctgamtilde$ and $\langle \tauhat \rangle|_{\gamma}$ for each
configuration keeping $\gamma$ constant and average then over all configurations, Eq.~(\ref{eq_CttauCtgamOne}).
Confirming Eq.~(\ref{eq_GtCttau}) we observe $G(t) \approx \Cttau$.

\paragraph*{Keeping the shear-barostat switched on.}
As shown by the small filled triangles in Fig.~\ref{fig_dataMD}, 
the same result is also obtained by sampling $\tauhat(t)$ 
for every $\delta t = 10^{-3}$ using an extremely slow shear-barostat for one single trajectory 
up to $t=10^5$. This large time is needed for a sufficient ensemble sampling.
Otherwise, $\muFttau$ would remain below its asymptotic large-$t$ limit $\muA=\muFtau$ \cite{foot_tdependence}. 
We have used here a hyprid MD-MC scheme where after every MD
step a Metropolis MC attempt was made to change the metric \cite{WXP13,foot_affineshear}
by a small amount $|\delta \gamma| \le \gammamax = 10^{-7}$.
Additional (non-universal) relaxation pathways become important if $\gamma(t)$ changes too strongly. 
If instead $\gammamax = 10^{-5}$ is used
(all other parameters kept constant) this naturally leads to a rapid decay of $\Cttau$ (crosses).
The conceptionally important point is here that in the limit of sufficiently small $\gammamax$
Eq.~(\ref{eq_weakbaro}) holds. The quenched $\NVtT$-ensemble average $\Cttau$ (open triangles) 
is then obtained without completely switching off the shear-barostat. 
The detailed description of the presumably non-universal
scaling of the additional relaxation pathways at larger $\gammamax$ is of course also of interest.
This should be addressed in future work.

\begin{figure}[t]
\centerline{\resizebox{1.0\columnwidth}{!}{\includegraphics*{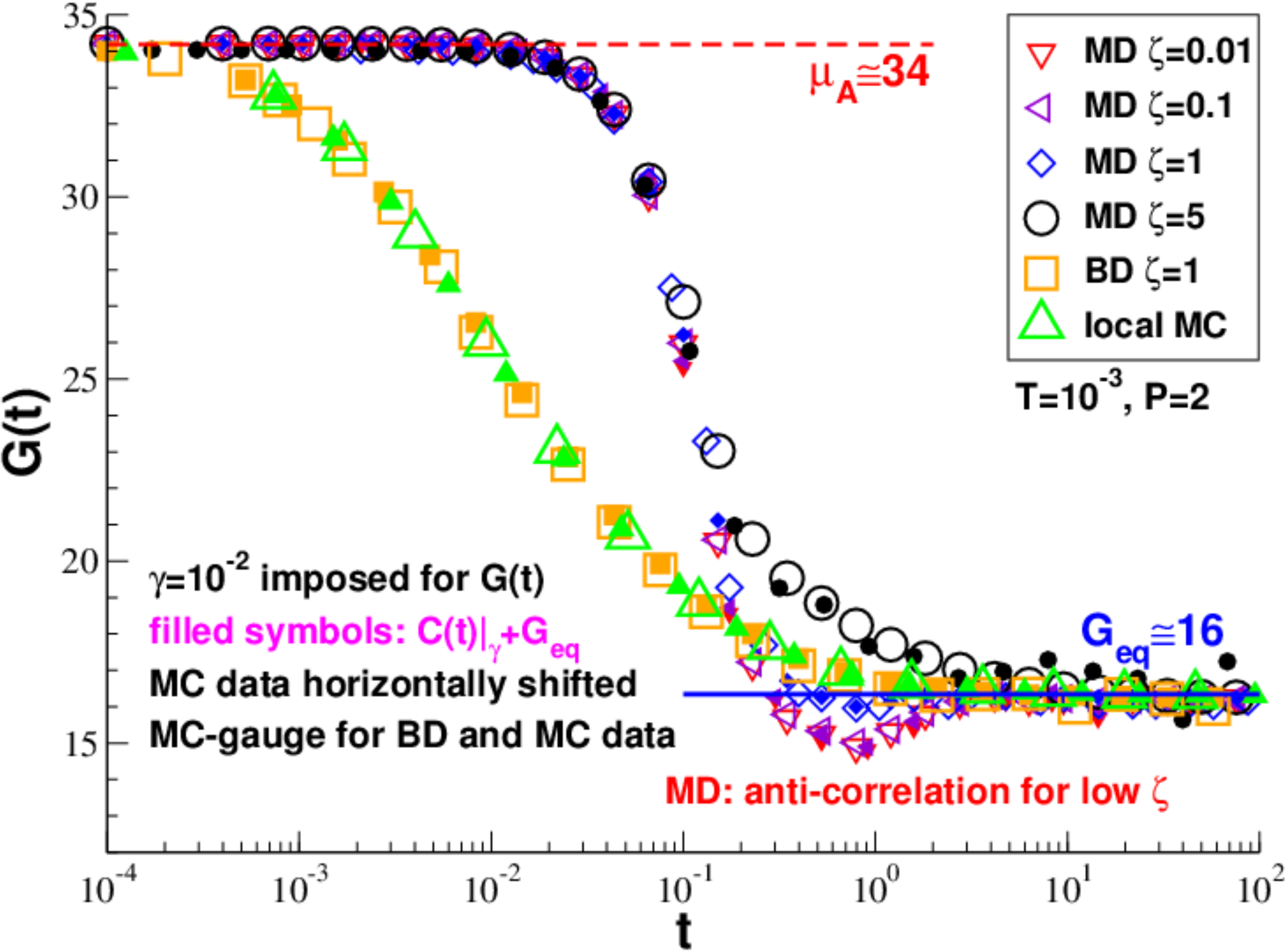}}}
\caption{Stress relaxation modulus $G(t)$ (open symbols) and rescaled autocorrelation function $\Ctgam+\Geq$
(filled symbols) for MD simulations for several friction constants $\zeta$, BD simulations
with $\zeta=1$ and local MC moves. 
Shifting horizontally the MC data by a factor $1/8000$ allows to bring them to collapse onto the BD data.
\label{fig_dataMDBDMC}
}
\end{figure}

\paragraph*{Different numerical schemes.}
The scaling collapse of $G(t)$ and $\Ctgam+\Geq$ has been also obtained for different temperatures $T$ 
(not shown) and friction constants $\zeta$ as may be seen from Fig.~\ref{fig_dataMDBDMC}. As one expects, 
the MD data decay more rapidly with decreasing $\zeta$ and reveal anti-correlations and oscillations
for the lowest $\zeta$ probed. Also included in Fig.~\ref{fig_dataMDBDMC} is data obtained by 
(overdamped) BD simulations with a friction constant $\zeta=1$ and MC simulations with 
local monomer jump attempts uniformally distributed in a disk of radius $0.01$ \cite{AllenTildesleyBook}.
Both data sets for each simulation type are again found to collapse. 
Note that it is possible to collapse additionally the BD and MC data by
shifting the MC data horizontally.

\paragraph*{Gauge freedom for the instantaneous stress.}
A technical point should finally be mentioned.
While for our MD simulations the instantaneous shear stress $\tauhat = \tauidhat + \tauexhat$
comprises both an ideal contribution $\tauidhat$
and an excess contribution $\tauexhat$
and correspondingly $\muA = \muAid + \muAex$ and $\muFgam = \muFid|_{\gamma} + \muFex|_{\gamma}$ take ideal contributions 
$\muAid = \muFid|_{\gamma} = \Pid$, this is not possible for BD and MC simulations. 
(Note that for the low temperature considered contributions of order $\Pid \approx 10^{-3}$ are in any case negligible.)
Within the so-called ``MC-gauge" \cite{WXP13c}, we thus set $\tauidhat \equiv 0$ for BD and MC, 
i.e. the kinetic degrees of freedom are considered to be integrated out.
Essentially we take advantage here of the general gauge freedom for the definition of instantaneous 
intensive variables \cite{foot_Ihat}. 
Note that for the demonstration of Eq.~(\ref{eq_key}) we did not specify whether the state 
of the system is characterized by the positions and momenta of the particles (as in MD simulations) 
or only by their positions (as in BD and MC).
To satisfy Eq.~(\ref{eq_key}) it is just required that $\muA$, $\muFgam$, $\Cttau$, $\Ctgam$ and $G(t)$ 
are {\em measured consistently}.

\section{Conclusion}
\label{sec_conc}
 
\paragraph*{Main results.}
Focusing on permanent isotropic networks in thermal equilibrium (Sec.~\ref{sec_model}) 
we have revisited theoretically (Sec.~\ref{theo_dyna}) and numerically (Sec.~\ref{simu_dyna})
the linear-response relation between the shear stress relaxation modulus $G(t)$ and 
the shear stress autocorrelation functions $\Ctgam$ and $\Cttau$ computed, respectively, 
at imposed strain $\gamma$ and mean stress $\tau$.
It has been demonstrated that according to Eq.~(\ref{eq_key}) or Eq.~(\ref{eq_Cttauctgam})
$G(t)=\Cttau$ and $\Ctgam$ must become different in the solid limit for $\Geq > 0$. 
While $G(t)$ may be determined numerically directly from $\Cttau$ using
either a quenched $\NVtT$-ensemble or an asymptotically slow shear-barostat
for which Eq.~(\ref{eq_weakbaro}) holds (Fig.~\ref{fig_dataMD}), 
this is not possible alone from $\Ctgam$. 

\paragraph*{Digression.}
More briefly, we have commented on self-assembled transient elastic networks 
characterized by a scission-recombination frequency $f$ for the springs (Sec.~\ref{theo_trans}).
For a finite, but small frequency $f$ the shear modulus $\Geq(f)$ must vanish for 
long sampling times. Following Ref.~\cite{Biroli10} we have argued that 
$G(t) = \Cttau = \Ctgam = \GFt$ should reveal an intermediate plateau $\GM$ 
and that this plateau is set by the finite shear modulus of the quenched network, $\GM = \Geq(f=0)$.

\paragraph*{Discussion.}
More generally, it is obviously often helpful to describe an observed intermediate plateau 
or shoulder of $G(t)$ in terms of a phenomenological shear modulus $\GM$ of a dynamical model, 
such as the Maxwell model for viscoelastic fluids or the reptation model of entangled polymer 
melts \cite{RubinsteinBook,DoiEdwardsBook}.
However, such a model allowing the theoretical {\em interpretation} of the data should not 
be confused with the proper {\em measurement procedure} and 
the model parameter $\GM$ should not be identified with the thermodynamic equilibrium modulus 
$\Geq$ of the system.
Note that the shear modulus $\Geq$ of a Maxwell fluid or a linear polymer melt must vanish
while the phenomenological parameter $\GM$ describing the short or intermediate time
stress response is finite.
Since in this sense different operational ``static" and ``dynamical" definitions of the shear modulus 
are used for describing glass-forming liquids close to the glass transition 
\cite{Klix12,Szamel11,Yoshino12,Mezard10,ZT13}, 
this may explain why qualitatively different temperature dependences
(cusp singularity \cite{Mezard10,ZT13} {\em vs.} finite jump \cite{GoetzeBook,Klix12,Szamel11,Ikeda12})
have been predicted. %
Hence, while our recent attempts to determine $\Geq(T)$ for two glass-forming model 
systems \cite{WXP13} are consistent with a continuous cusp, this is not necessarily 
in contradiction with a jump singularity for $\GM(T)$ determined from $\Ctgam$ \cite{Klix12,Ikeda12}.

\paragraph*{Outlook.}
It should be noted that generalizing Eq.~(\ref{eq_key}) one obtains readily that
\begin{equation}
M(t) = C(t)|_I = C(t)|_{X} + \Meq \ \mbox{ for }  t >0
\label{eq_final}
\end{equation}
with $M(t)$ being the relaxation modulus 
of an intensive variable $I$, $\Meq = \partial I / \partial X$ the associated equilibrium modulus
and $C(t) = \beta V \langle \delta \Ihat(t) \delta \Ihat(0) \rangle$
the corresponding autocorrelation function for any (continuous) {\em intensive} variable $\Ihat(t)$.
We note finally that we are currently simulating transient elastic networks formed by dense, purely repulsive beads 
which are reversibly connected by harmonic springs. The preliminary results support Eq.~(\ref{eq_transnet})
suggested in Sec.~\ref{theo_trans}, i.e. it is seen that $G(t)$, $\Ctgam$ and $\GFt$ 
approach with decreasing, but finite scission-recombination frequency $f$, i.e. $\Geq(f)=0$,
an intermediate plateau given by the shear modulus $\Geq(f=0)$ of the quenched reference network \cite{foot_muF}.

\vspace*{0.2cm} 
\begin{acknowledgments}
H.X. thanks the IRTG Soft Matter for financial support.
We are indebted to O.~Benzerara and J.~Farago (both ICS, Strasbourg) and
M. Fuchs (Konstanz) for helpful discussions.
\end{acknowledgments}


\begin{thebibliography}{49}
\expandafter\ifx\csname natexlab\endcsname\relax\def\natexlab#1{#1}\fi
\expandafter\ifx\csname bibnamefont\endcsname\relax
  \def\bibnamefont#1{#1}\fi
\expandafter\ifx\csname bibfnamefont\endcsname\relax
  \def\bibfnamefont#1{#1}\fi
\expandafter\ifx\csname citenamefont\endcsname\relax
  \def\citenamefont#1{#1}\fi
\expandafter\ifx\csname url\endcsname\relax
  \def\url#1{\texttt{#1}}\fi
\expandafter\ifx\csname urlprefix\endcsname\relax\def\urlprefix{URL }\fi
\providecommand{\bibinfo}[2]{#2}
\providecommand{\eprint}[2][]{\url{#2}}

\bibitem[{\citenamefont{Landau and Lifshitz}(1959)}]{LandauElasticity}
\bibinfo{author}{\bibfnamefont{L.~D.} \bibnamefont{Landau}} \bibnamefont{and}
  \bibinfo{author}{\bibfnamefont{E.~M.} \bibnamefont{Lifshitz}},
  \emph{\bibinfo{title}{Theory of Elasticity}} (\bibinfo{publisher}{Pergamon
  Press}, \bibinfo{year}{1959}).

\bibitem[{\citenamefont{Rubinstein and Colby}(2003)}]{RubinsteinBook}
\bibinfo{author}{\bibfnamefont{M.}~\bibnamefont{Rubinstein}} \bibnamefont{and}
  \bibinfo{author}{\bibfnamefont{R.}~\bibnamefont{Colby}},
  \emph{\bibinfo{title}{Polymer Physics}} (\bibinfo{publisher}{Oxford
  University Press}, \bibinfo{address}{Oxford}, \bibinfo{year}{2003}).

\bibitem[{\citenamefont{Doi and Edwards}(1986)}]{DoiEdwardsBook}
\bibinfo{author}{\bibfnamefont{M.}~\bibnamefont{Doi}} \bibnamefont{and}
  \bibinfo{author}{\bibfnamefont{S.~F.} \bibnamefont{Edwards}},
  \emph{\bibinfo{title}{The Theory of Polymer Dynamics}}
  (\bibinfo{publisher}{Clarendon Press}, \bibinfo{address}{Oxford},
  \bibinfo{year}{1986}).

\bibitem[{\citenamefont{Hansen and McDonald}(2006)}]{HansenBook}
\bibinfo{author}{\bibfnamefont{J.}~\bibnamefont{Hansen}} \bibnamefont{and}
  \bibinfo{author}{\bibfnamefont{I.}~\bibnamefont{McDonald}},
  \emph{\bibinfo{title}{Theory of simple liquids}}
  (\bibinfo{publisher}{Academic Press}, \bibinfo{address}{New York},
  \bibinfo{year}{2006}), \bibinfo{note}{3nd edition}.

\bibitem[{\citenamefont{G\"otze}(2009)}]{GoetzeBook}
\bibinfo{author}{\bibfnamefont{W.}~\bibnamefont{G\"otze}},
  \emph{\bibinfo{title}{Complex Dynamics of Glass-Forming Liquids: A
  Mode-Coupling Theory}} (\bibinfo{publisher}{Oxford University Press, Oxford},
  \bibinfo{year}{2009}).

\bibitem[{\citenamefont{Alexander}(1998)}]{Alexander98}
\bibinfo{author}{\bibfnamefont{S.}~\bibnamefont{Alexander}},
  \bibinfo{journal}{Physics Reports} \textbf{\bibinfo{volume}{296}},
  \bibinfo{pages}{65 } (\bibinfo{year}{1998}).

\bibitem[{\citenamefont{Witten and Pincus}(2004)}]{WittenPincusBook}
\bibinfo{author}{\bibfnamefont{T.}~\bibnamefont{Witten}} \bibnamefont{and}
  \bibinfo{author}{\bibfnamefont{P.~A.} \bibnamefont{Pincus}},
  \emph{\bibinfo{title}{Structured Fluids: Polymers, Colloids, Surfactants}}
  (\bibinfo{publisher}{Oxford University Press}, \bibinfo{address}{Oxford},
  \bibinfo{year}{2004}).

\bibitem[{\citenamefont{Callen}(1985)}]{Callen}
\bibinfo{author}{\bibfnamefont{H.~B.} \bibnamefont{Callen}},
  \emph{\bibinfo{title}{Thermodynamics and an Introduction to
  Thermostatistics}} (\bibinfo{publisher}{Wiley}, \bibinfo{address}{New York},
  \bibinfo{year}{1985}).

\bibitem[{\citenamefont{Chandler}(1987)}]{ChandlerBook}
\bibinfo{author}{\bibfnamefont{D.}~\bibnamefont{Chandler}},
  \emph{\bibinfo{title}{Introduction to Modern Statistical Mechanics}}
  (\bibinfo{publisher}{Oxford University Press}, \bibinfo{address}{New York},
  \bibinfo{year}{1987}).

\bibitem[{\citenamefont{Chaikin and Lubensky}(1995)}]{ChaikinBook}
\bibinfo{author}{\bibfnamefont{P.~M.} \bibnamefont{Chaikin}} \bibnamefont{and}
  \bibinfo{author}{\bibfnamefont{T.~C.} \bibnamefont{Lubensky}},
  \emph{\bibinfo{title}{Principles of condensed matter physics}}
  (\bibinfo{publisher}{Cambridge University Press}, \bibinfo{year}{1995}).

\bibitem[{\citenamefont{Sausset et~al.}(2010)\citenamefont{Sausset, Biroli, and
  Kurchan}}]{Biroli10}
\bibinfo{author}{\bibfnamefont{F.}~\bibnamefont{Sausset}},
  \bibinfo{author}{\bibfnamefont{G.}~\bibnamefont{Biroli}}, \bibnamefont{and}
  \bibinfo{author}{\bibfnamefont{J.}~\bibnamefont{Kurchan}},
  \bibinfo{journal}{J. Stat. Phys.} \textbf{\bibinfo{volume}{140}},
  \bibinfo{pages}{718} (\bibinfo{year}{2010}).

\bibitem[{\citenamefont{Barrat et~al.}(1988)\citenamefont{Barrat, Roux, Hansen,
  and Klein}}]{Barrat88}
\bibinfo{author}{\bibfnamefont{J.-L.} \bibnamefont{Barrat}},
  \bibinfo{author}{\bibfnamefont{J.-N.} \bibnamefont{Roux}},
  \bibinfo{author}{\bibfnamefont{J.-P.} \bibnamefont{Hansen}},
  \bibnamefont{and} \bibinfo{author}{\bibfnamefont{M.~L.} \bibnamefont{Klein}},
  \bibinfo{journal}{Europhys. Lett.} \textbf{\bibinfo{volume}{7}},
  \bibinfo{pages}{707} (\bibinfo{year}{1988}).

\bibitem[{\citenamefont{Wittmer et~al.}(2002)\citenamefont{Wittmer, Tanguy,
  Barrat, and Lewis}}]{WTBL02}
\bibinfo{author}{\bibfnamefont{J.~P.} \bibnamefont{Wittmer}},
  \bibinfo{author}{\bibfnamefont{A.}~\bibnamefont{Tanguy}},
  \bibinfo{author}{\bibfnamefont{J.-L.} \bibnamefont{Barrat}},
  \bibnamefont{and} \bibinfo{author}{\bibfnamefont{L.}~\bibnamefont{Lewis}},
  \bibinfo{journal}{Europhys. Lett.} \textbf{\bibinfo{volume}{57}},
  \bibinfo{pages}{423} (\bibinfo{year}{2002}).

\bibitem[{\citenamefont{Tanguy et~al.}(2002)\citenamefont{Tanguy, Wittmer,
  Leonforte, and Barrat}}]{TWLB02}
\bibinfo{author}{\bibfnamefont{A.}~\bibnamefont{Tanguy}},
  \bibinfo{author}{\bibfnamefont{J.~P.} \bibnamefont{Wittmer}},
  \bibinfo{author}{\bibfnamefont{F.}~\bibnamefont{Leonforte}},
  \bibnamefont{and} \bibinfo{author}{\bibfnamefont{J.-L.}
  \bibnamefont{Barrat}}, \bibinfo{journal}{Phys. Rev. B}
  \textbf{\bibinfo{volume}{66}}, \bibinfo{pages}{174205}
  (\bibinfo{year}{2002}).

\bibitem[{\citenamefont{Berthier et~al.}(2005)\citenamefont{Berthier, Biroli,
  Bouchaud, Cipelletti, Masri, L'H\^ote, Ladieu, and Pierno}}]{Berthier05}
\bibinfo{author}{\bibfnamefont{L.}~\bibnamefont{Berthier}},
  \bibinfo{author}{\bibfnamefont{G.}~\bibnamefont{Biroli}},
  \bibinfo{author}{\bibfnamefont{J.-P.} \bibnamefont{Bouchaud}},
  \bibinfo{author}{\bibfnamefont{L.}~\bibnamefont{Cipelletti}},
  \bibinfo{author}{\bibfnamefont{D.~E.} \bibnamefont{Masri}},
  \bibinfo{author}{\bibfnamefont{D.}~\bibnamefont{L'H\^ote}},
  \bibinfo{author}{\bibfnamefont{F.}~\bibnamefont{Ladieu}}, \bibnamefont{and}
  \bibinfo{author}{\bibfnamefont{M.}~\bibnamefont{Pierno}},
  \bibinfo{journal}{Phys. Rev. Lett.} \textbf{\bibinfo{volume}{310}},
  \bibinfo{pages}{1797} (\bibinfo{year}{2005}).

\bibitem[{\citenamefont{Berthier et~al.}(2007)\citenamefont{Berthier, Biroli,
  Bouchaud, Kob, Miyazaki, and Reichman}}]{Berthier07}
\bibinfo{author}{\bibfnamefont{L.}~\bibnamefont{Berthier}},
  \bibinfo{author}{\bibfnamefont{G.}~\bibnamefont{Biroli}},
  \bibinfo{author}{\bibfnamefont{J.-P.} \bibnamefont{Bouchaud}},
  \bibinfo{author}{\bibfnamefont{W.}~\bibnamefont{Kob}},
  \bibinfo{author}{\bibfnamefont{K.}~\bibnamefont{Miyazaki}}, \bibnamefont{and}
  \bibinfo{author}{\bibfnamefont{D.}~\bibnamefont{Reichman}},
  \bibinfo{journal}{J. Chem. Phys.} \textbf{\bibinfo{volume}{126}},
  \bibinfo{pages}{184503} (\bibinfo{year}{2007}).

\bibitem[{\citenamefont{Yoshino and M\'ezard}(2010)}]{Mezard10}
\bibinfo{author}{\bibfnamefont{H.}~\bibnamefont{Yoshino}} \bibnamefont{and}
  \bibinfo{author}{\bibfnamefont{M.}~\bibnamefont{M\'ezard}},
  \bibinfo{journal}{Phys. Rev. Lett.} \textbf{\bibinfo{volume}{105}},
  \bibinfo{pages}{015504} (\bibinfo{year}{2010}).

\bibitem[{\citenamefont{Szamel and Flenner}(2011)}]{Szamel11}
\bibinfo{author}{\bibfnamefont{G.}~\bibnamefont{Szamel}} \bibnamefont{and}
  \bibinfo{author}{\bibfnamefont{E.}~\bibnamefont{Flenner}},
  \bibinfo{journal}{Phys. Rev. Lett.} \textbf{\bibinfo{volume}{107}},
  \bibinfo{pages}{105505} (\bibinfo{year}{2011}).

\bibitem[{\citenamefont{Schnell et~al.}(2011)\citenamefont{Schnell, Meyer,
  Fond, Wittmer, and Baschnagel}}]{SBM11}
\bibinfo{author}{\bibfnamefont{B.}~\bibnamefont{Schnell}},
  \bibinfo{author}{\bibfnamefont{H.}~\bibnamefont{Meyer}},
  \bibinfo{author}{\bibfnamefont{C.}~\bibnamefont{Fond}},
  \bibinfo{author}{\bibfnamefont{J.~P.} \bibnamefont{Wittmer}},
  \bibnamefont{and}
  \bibinfo{author}{\bibfnamefont{J.}~\bibnamefont{Baschnagel}},
  \bibinfo{journal}{Eur. Phys. J. E} \textbf{\bibinfo{volume}{34}},
  \bibinfo{pages}{97} (\bibinfo{year}{2011}).

\bibitem[{\citenamefont{Yoshino}(2012)}]{Yoshino12}
\bibinfo{author}{\bibfnamefont{H.}~\bibnamefont{Yoshino}}, \bibinfo{journal}{J.
  Chem. Phys.} \textbf{\bibinfo{volume}{136}}, \bibinfo{pages}{214108}
  (\bibinfo{year}{2012}).

\bibitem[{\citenamefont{Klix et~al.}(2012)\citenamefont{Klix, Ebert, Weysser,
  Fuchs, Maret, and Keim}}]{Klix12}
\bibinfo{author}{\bibfnamefont{C.}~\bibnamefont{Klix}},
  \bibinfo{author}{\bibfnamefont{F.}~\bibnamefont{Ebert}},
  \bibinfo{author}{\bibfnamefont{F.}~\bibnamefont{Weysser}},
  \bibinfo{author}{\bibfnamefont{M.}~\bibnamefont{Fuchs}},
  \bibinfo{author}{\bibfnamefont{G.}~\bibnamefont{Maret}}, \bibnamefont{and}
  \bibinfo{author}{\bibfnamefont{P.}~\bibnamefont{Keim}},
  \bibinfo{journal}{Phys. Rev. Lett.} \textbf{\bibinfo{volume}{109}},
  \bibinfo{pages}{178301} (\bibinfo{year}{2012}).

\bibitem[{\citenamefont{Xu et~al.}(2012)\citenamefont{Xu, Wittmer,
  Poli{\'n}ska, and Baschnagel}}]{XWP12}
\bibinfo{author}{\bibfnamefont{H.}~\bibnamefont{Xu}},
  \bibinfo{author}{\bibfnamefont{J.}~\bibnamefont{Wittmer}},
  \bibinfo{author}{\bibfnamefont{P.}~\bibnamefont{Poli{\'n}ska}},
  \bibnamefont{and}
  \bibinfo{author}{\bibfnamefont{J.}~\bibnamefont{Baschnagel}},
  \bibinfo{journal}{Phys. Rev. E} \textbf{\bibinfo{volume}{86}},
  \bibinfo{pages}{046705} (\bibinfo{year}{2012}).

\bibitem[{\citenamefont{Wittmer
  et~al.}(2013{\natexlab{a}})\citenamefont{Wittmer, Xu, Poli\'nska, Weysser,
  and Baschnagel}}]{WXP13}
\bibinfo{author}{\bibfnamefont{J.~P.} \bibnamefont{Wittmer}},
  \bibinfo{author}{\bibfnamefont{H.}~\bibnamefont{Xu}},
  \bibinfo{author}{\bibfnamefont{P.}~\bibnamefont{Poli\'nska}},
  \bibinfo{author}{\bibfnamefont{F.}~\bibnamefont{Weysser}}, \bibnamefont{and}
  \bibinfo{author}{\bibfnamefont{J.}~\bibnamefont{Baschnagel}},
  \bibinfo{journal}{J. Chem. Phys.} \textbf{\bibinfo{volume}{138}},
  \bibinfo{pages}{12A533} (\bibinfo{year}{2013}{\natexlab{a}}).

\bibitem[{\citenamefont{Wittmer
  et~al.}(2013{\natexlab{b}})\citenamefont{Wittmer, Xu, Poli\'nska, Gillig,
  Hellferich, Weysser, and Baschnagel}}]{WXP13c}
\bibinfo{author}{\bibfnamefont{J.~P.} \bibnamefont{Wittmer}},
  \bibinfo{author}{\bibfnamefont{H.}~\bibnamefont{Xu}},
  \bibinfo{author}{\bibfnamefont{P.}~\bibnamefont{Poli\'nska}},
  \bibinfo{author}{\bibfnamefont{C.}~\bibnamefont{Gillig}},
  \bibinfo{author}{\bibfnamefont{J.}~\bibnamefont{Hellferich}},
  \bibinfo{author}{\bibfnamefont{F.}~\bibnamefont{Weysser}}, \bibnamefont{and}
  \bibinfo{author}{\bibfnamefont{J.}~\bibnamefont{Baschnagel}},
  \bibinfo{journal}{Eur. Phys. J. E} \textbf{\bibinfo{volume}{36}},
  \bibinfo{pages}{131} (\bibinfo{year}{2013}{\natexlab{b}}).

\bibitem[{\citenamefont{Zaccone and Terentjev}(2013)}]{ZT13}
\bibinfo{author}{\bibfnamefont{A.}~\bibnamefont{Zaccone}} \bibnamefont{and}
  \bibinfo{author}{\bibfnamefont{E.}~\bibnamefont{Terentjev}},
  \bibinfo{journal}{Phys. Rev. Lett.} \textbf{\bibinfo{volume}{110}},
  \bibinfo{pages}{178002} (\bibinfo{year}{2013}).

\bibitem[{\citenamefont{Ozawa et~al.}(2012)\citenamefont{Ozawa, Kuroiwa, and
  Ikeda}}]{Ikeda12}
\bibinfo{author}{\bibfnamefont{M.}~\bibnamefont{Ozawa}},
  \bibinfo{author}{\bibfnamefont{T.}~\bibnamefont{Kuroiwa}}, \bibnamefont{and}
  \bibinfo{author}{\bibfnamefont{A.}~\bibnamefont{Ikeda}},
  \bibinfo{journal}{Phys. Rev. Lett.} \textbf{\bibinfo{volume}{109}},
  \bibinfo{pages}{205701} (\bibinfo{year}{2012}).

\bibitem[{\citenamefont{Mizuno et~al.}(2013)\citenamefont{Mizuno, Mossa, and
  Barrat}}]{Barrat13}
\bibinfo{author}{\bibfnamefont{H.}~\bibnamefont{Mizuno}},
  \bibinfo{author}{\bibfnamefont{S.}~\bibnamefont{Mossa}}, \bibnamefont{and}
  \bibinfo{author}{\bibfnamefont{J.-L.} \bibnamefont{Barrat}},
  \bibinfo{journal}{Phys. Rev. E} \textbf{\bibinfo{volume}{87}},
  \bibinfo{pages}{042306} (\bibinfo{year}{2013}).

\bibitem[{\citenamefont{del Gado and Kob}(2008)}]{Kob08}
\bibinfo{author}{\bibfnamefont{E.}~\bibnamefont{del Gado}} \bibnamefont{and}
  \bibinfo{author}{\bibfnamefont{W.}~\bibnamefont{Kob}}, \bibinfo{journal}{J.
  Non-Newtonian Fluid Mech.} \textbf{\bibinfo{volume}{149}},
  \bibinfo{pages}{28} (\bibinfo{year}{2008}).

\bibitem[{\citenamefont{Duering et~al.}(1991)\citenamefont{Duering, Kremer, and
  Grest}}]{DKG91}
\bibinfo{author}{\bibfnamefont{E.}~\bibnamefont{Duering}},
  \bibinfo{author}{\bibfnamefont{K.}~\bibnamefont{Kremer}}, \bibnamefont{and}
  \bibinfo{author}{\bibfnamefont{G.~S.} \bibnamefont{Grest}},
  \bibinfo{journal}{Phys. Rev. Lett.} \textbf{\bibinfo{volume}{3531}},
  \bibinfo{pages}{67} (\bibinfo{year}{1991}).

\bibitem[{\citenamefont{Duering et~al.}(1994)\citenamefont{Duering, Kremer, and
  Grest}}]{DKG94}
\bibinfo{author}{\bibfnamefont{E.}~\bibnamefont{Duering}},
  \bibinfo{author}{\bibfnamefont{K.}~\bibnamefont{Kremer}}, \bibnamefont{and}
  \bibinfo{author}{\bibfnamefont{G.~S.} \bibnamefont{Grest}},
  \bibinfo{journal}{J. Chem. Phys.} \textbf{\bibinfo{volume}{8169}},
  \bibinfo{pages}{101} (\bibinfo{year}{1994}).

\bibitem[{\citenamefont{Ulrich et~al.}(2006)\citenamefont{Ulrich, Mao,
  Goldbart, and Zippelius}}]{Zippelius06}
\bibinfo{author}{\bibfnamefont{S.}~\bibnamefont{Ulrich}},
  \bibinfo{author}{\bibfnamefont{X.}~\bibnamefont{Mao}},
  \bibinfo{author}{\bibfnamefont{P.}~\bibnamefont{Goldbart}}, \bibnamefont{and}
  \bibinfo{author}{\bibfnamefont{A.}~\bibnamefont{Zippelius}},
  \bibinfo{journal}{Europhysics Lett.} \textbf{\bibinfo{volume}{76}},
  \bibinfo{pages}{677} (\bibinfo{year}{2006}).

\bibitem[{\citenamefont{Tonhauser et~al.}(2010)\citenamefont{Tonhauser, Wilms,
  Korth, Frey, and Friedrich}}]{Friedrich10}
\bibinfo{author}{\bibfnamefont{C.}~\bibnamefont{Tonhauser}},
  \bibinfo{author}{\bibfnamefont{D.}~\bibnamefont{Wilms}},
  \bibinfo{author}{\bibfnamefont{Y.}~\bibnamefont{Korth}},
  \bibinfo{author}{\bibfnamefont{H.}~\bibnamefont{Frey}}, \bibnamefont{and}
  \bibinfo{author}{\bibfnamefont{C.}~\bibnamefont{Friedrich}},
  \bibinfo{journal}{Macromolecular Rapid Comm.} \textbf{\bibinfo{volume}{31}},
  \bibinfo{pages}{2127} (\bibinfo{year}{2010}).

\bibitem[{\citenamefont{Zilman et~al.}(2003)\citenamefont{Zilman, Kieffer,
  Molino, Porte, and Safran}}]{Porte03}
\bibinfo{author}{\bibfnamefont{A.}~\bibnamefont{Zilman}},
  \bibinfo{author}{\bibfnamefont{J.}~\bibnamefont{Kieffer}},
  \bibinfo{author}{\bibfnamefont{F.}~\bibnamefont{Molino}},
  \bibinfo{author}{\bibfnamefont{G.}~\bibnamefont{Porte}}, \bibnamefont{and}
  \bibinfo{author}{\bibfnamefont{S.~A.} \bibnamefont{Safran}},
  \bibinfo{journal}{Phys. Rev. Lett.} \textbf{\bibinfo{volume}{91}},
  \bibinfo{pages}{2003} (\bibinfo{year}{2003}).

\bibitem[{foo({\natexlab{a}})}]{foot_EOS}
\bibinfo{note}{Numerically, it might be better to measure the average shear
  stress $\tau(\gamma)$, to fit a spline to this equation of state and to
  determine $\Geq(\gamma)$ from the fit by taking the derivative with respect
  to the shear strain $\gamma$.}

\bibitem[{\citenamefont{Parrinello and Rahman}(1982)}]{Parinello82}
\bibinfo{author}{\bibfnamefont{M.}~\bibnamefont{Parrinello}} \bibnamefont{and}
  \bibinfo{author}{\bibfnamefont{A.}~\bibnamefont{Rahman}},
  \bibinfo{journal}{J. Chem. Phys.} \textbf{\bibinfo{volume}{76}},
  \bibinfo{pages}{2662} (\bibinfo{year}{1982}).

\bibitem[{\citenamefont{Lutsko}(1989)}]{Lutsko89}
\bibinfo{author}{\bibfnamefont{J.~F.} \bibnamefont{Lutsko}},
  \bibinfo{journal}{J. Appl. Phys} \textbf{\bibinfo{volume}{65}},
  \bibinfo{pages}{2991} (\bibinfo{year}{1989}).

\bibitem[{\citenamefont{Allen and Tildesley}(1994)}]{AllenTildesleyBook}
\bibinfo{author}{\bibfnamefont{M.}~\bibnamefont{Allen}} \bibnamefont{and}
  \bibinfo{author}{\bibfnamefont{D.}~\bibnamefont{Tildesley}},
  \emph{\bibinfo{title}{Computer Simulation of Liquids}}
  (\bibinfo{publisher}{Oxford University Press}, \bibinfo{address}{Oxford},
  \bibinfo{year}{1994}).

\bibitem[{\citenamefont{Frenkel and Smit}(2002)}]{FrenkelSmitBook}
\bibinfo{author}{\bibfnamefont{D.}~\bibnamefont{Frenkel}} \bibnamefont{and}
  \bibinfo{author}{\bibfnamefont{B.}~\bibnamefont{Smit}},
  \emph{\bibinfo{title}{Understanding Molecular Simulation -- From Algorithms
  to Applications}} (\bibinfo{publisher}{Academic Press}, \bibinfo{address}{San
  Diego}, \bibinfo{year}{2002}), \bibinfo{note}{2nd edition}.

\bibitem[{\citenamefont{Thijssen}(1999)}]{ThijssenBook}
\bibinfo{author}{\bibfnamefont{J.}~\bibnamefont{Thijssen}},
  \emph{\bibinfo{title}{Computational Physics}} (\bibinfo{publisher}{Cambridge
  University Press}, \bibinfo{address}{Cambridge}, \bibinfo{year}{1999}).

\bibitem[{\citenamefont{Landau and Binder}(2000)}]{LandauBinderBook}
\bibinfo{author}{\bibfnamefont{D.~P.} \bibnamefont{Landau}} \bibnamefont{and}
  \bibinfo{author}{\bibfnamefont{K.}~\bibnamefont{Binder}},
  \emph{\bibinfo{title}{A Guide to Monte Carlo Simulations in Statistical
  Physics}} (\bibinfo{publisher}{Cambridge University Press},
  \bibinfo{address}{Cambridge}, \bibinfo{year}{2000}).

\bibitem[{foo({\natexlab{b}})}]{foot_affineshear}
\bibinfo{note}{A canonical affine transformation consists in changing both the
  particle coordinates $\rvec \Rightarrow \hmat \ \rvec$ and the conjugated
  velocities $\vvec \to \hmat^{-1} \vvec$ using a linear matrix $\hmat$
  \cite{Parinello82,Lutsko89}. For a uniform shear in two dimensions this
  amounts to $\rx \to \rx + \gamma \ry$ and $\vx \to \vx - \gamma \vy$ for the
  transformation of the $x$-components of the particle positions and
  velocities.}

\bibitem[{\citenamefont{Green}(1960)}]{Green60}
\bibinfo{author}{\bibfnamefont{M.~S.} \bibnamefont{Green}},
  \bibinfo{journal}{Phys. Rev.} \textbf{\bibinfo{volume}{119}},
  \bibinfo{pages}{829} (\bibinfo{year}{1960}).

\bibitem[{\citenamefont{Lebowitz et~al.}(1967)\citenamefont{Lebowitz, Percus,
  and Verlet}}]{Lebowitz67}
\bibinfo{author}{\bibfnamefont{J.~L.} \bibnamefont{Lebowitz}},
  \bibinfo{author}{\bibfnamefont{J.~K.} \bibnamefont{Percus}},
  \bibnamefont{and} \bibinfo{author}{\bibfnamefont{L.}~\bibnamefont{Verlet}},
  \bibinfo{journal}{Phys. Rev.} \textbf{\bibinfo{volume}{153}},
  \bibinfo{pages}{250} (\bibinfo{year}{1967}).

\bibitem[{foo({\natexlab{c}})}]{foot_tdependence}
\bibinfo{note}{The notation $\muF$, Eq.~(\ref{eq_muFdef}), or $\GF$,
  Eq.~(\ref{eq_GeqNVgT}), without time argument refer to static thermodynamic
  properties determined using asymptotically long sampling times. In order to
  avoid additional notations we write $\muF(t)$ or $\GFt$ to indicate that
  these properties have been determined using a finite time window $t$. Please
  note also that $\GFt$ should not be confused with the response function
  $G(t)$ albeit both properties are related as discussed in
  Sec.~\ref{theo_muFtCt}.}

\bibitem[{foo({\natexlab{d}})}]{foot_Ihat}
\bibinfo{note}{In general there is a considerable freedom for defining an {\em
  instantaneous} intensive variable $\Ihat$ as long as its average $I = \langle
  \Ihat \rangle$ does not change. The experimentally natural choice for
  $\Ihat=\tauhat$ is given by the instantaneous forces acting on the shear cell
  boundaries. For theory and simulation, $\tauhat \equiv \Hhat^{\prime}(\gamma
  V)$, i.e. the energy change with respect to an assumed {\em affine} strain,
  is the most common choice due to Eq.~(\ref{eq_pprime}).}

\bibitem[{foo({\natexlab{e}})}]{foot_Edwards}
\bibinfo{note}{It is instructive to obtain $G(t) = \Geq + \Ctgam$ directly by
  rewriting the derivation of the FDT given in Ref.~\cite{DoiEdwardsBook} for a
  strain $\gamma$ switched on at $t=0$. This shows that the $\Geq$-term stems
  naturally from the residual finite shear stress of the strained system at
  equilibrium.}

\bibitem[{foo({\natexlab{f}})}]{foot_Debye}
\bibinfo{note}{For an exponentially decaying stress-autocorrelation function
  $C(t) = C(0) \exp(-x)$ with $x = t/\tau$ one obtains by integration
  $\muF(t)/\muF = 1 - \fDebye(x)$ with $\fDebye(x) \equiv 2 (\exp(-x)
  -1+x)/x^2$ being the Debye function well-known in polymer science
  \cite{DoiEdwardsBook}. For large times $x \gg 1$ one thus obtains
  $\muF(t)/\muF = 1 - 2/x$.}

\bibitem[{\citenamefont{Wittmer et~al.}(1998)\citenamefont{Wittmer, Milchev,
  and Cates}}]{WMC98b}
\bibinfo{author}{\bibfnamefont{J.~P.} \bibnamefont{Wittmer}},
  \bibinfo{author}{\bibfnamefont{A.}~\bibnamefont{Milchev}}, \bibnamefont{and}
  \bibinfo{author}{\bibfnamefont{M.~E.} \bibnamefont{Cates}},
  \bibinfo{journal}{J. Chem. Phys.} \textbf{\bibinfo{volume}{109}},
  \bibinfo{pages}{834} (\bibinfo{year}{1998}).

\bibitem[{foo({\natexlab{g}})}]{foot_muF}
\bibinfo{note}{All simple means are independent of $f$, especially $\muA$.
  Interestingly, this is different for the shear stress fluctuations $\muFgam$,
  i.e. a well-defined static four-point correlation function, which shows a
  singular behavior for asymptotically long sampling times. While $\muFgam =
  \muA- \Geq(f=0)$ for the quenched network, one obtains $\muFgam = \muFtau =
  \muA$ for finite $f$ (consistent with $\Geq=0$).}

\end{thebibliography}

\end{document}